\documentclass[journal]{IEEEtran}
\usepackage{amsmath,amsfonts}
\usepackage{algorithmic}
\usepackage{algorithm}
\usepackage{array}
\usepackage[caption=false,font=normalsize,labelfont=sf,textfont=sf]{subfig}
\usepackage{textcomp}
\usepackage{stfloats}
\usepackage{url}
\usepackage{verbatim}
\usepackage{graphicx}
\usepackage{cite}

\usepackage{soul}

\usepackage[table,xcdraw]{xcolor}
\sodef\tightnum{}{-.1em}{0.1em}{-.1em}
\sodef\tightword{}{-.05em}{0.1em}{-.05em}

\begin{document}

\title{Implementing Homomorphic Encryption-Based Logic Locking in System-on-Chip Designs}

\author{Ye Ziyang,~\IEEEmembership{Student Member,~IEEE}, Makoto Ikeda,~\IEEEmembership{Senior Member,~IEEE}
\thanks{The authors are with the Department of Electrical Engineering and Information Systems, Graduate School of Engineering, The University of Tokyo, Tokyo, Japan (email: ye-zy@g.ecc.u-tokyo.ac.jp; ikeda@silicon.u-tokyo.ac.jp).}

}

\markboth{Journal of \LaTeX\ Class Files,~Vol.~14, No.~8, August~2021}%
{Shell \MakeLowercase{\textit{et al.}}: A Sample Article Using IEEEtran.cls for IEEE Journals}

\IEEEpubid{0000--0000/00\$00.00~\copyright~2021 IEEE}

\maketitle

\begin{abstract}
This study presents a logic locking scheme based on the binary Ring Learning With Errors algorithm, implemented in a RISC-V System-on-Chip design. Unlike traditional logic locking methods that require providing users with raw locking parameters, the proposed approach secures critical logic paths in the privilege switching process without exposing these sensitive parameters. The implemented locking module itself consumes 3519 Look-Up Tables and 2645 Registers, leading to an overall overhead of 6.0\% in Look-Up Tables and 6.9\% in Registers compared to the baseline system. The unlock process requires about 2.6\textmu s, introducing moderate performance impact, primarily affecting system-level operations while preserving user-level computational efficiency.
\end{abstract}


\section{Introduction}

\IEEEPARstart{H}{ardware} security has become increasingly vital in critical infrastructure and everyday life. The integrity of hardware components fundamentally influences the security of entire computing ecosystems, playing a crucial role in safeguarding sensitive data, personal privacy, and proprietary information. Currently, reverse engineering techniques \cite{rajarathnam2020regds} pose significant challenges to the integrity of digital circuits. Upon successful reverse engineering of a circuit, adversaries can execute malicious modifications, potentially compromising its intended functionality and security properties. Logic locking has emerged as a promising countermeasure to these vulnerabilities.

Traditional logic locking techniques, while widely adopted for hardware security, face two significant challenges. First, existing logic locking methods, such as \cite{rathor2024gatelock}, utilize identical parameters for both locking and unlocking operations, making them susceptible to parameter leakage and potentially exposing the original circuit design to malicious actors. Second, the emergence of quantum computing threatens the fundamental security assumptions of current logic locking methods, particularly those based on Boolean satisfiability (SAT) problems. Recent advances in quantum SAT algorithms \cite{boulebnane2024solving} have demonstrated the potential to reduce SAT problem complexity to $e^{0.7n}$, significantly undermining the computational security of traditional logic locking approaches. These challenges create an urgent need for more robust hardware security solutions that can withstand both contemporary and future threats.

To address these challenges, we propose a novel approach that integrates homomorphic encryption, specifically the binary Ring Learning With Errors (bin-RLWE) scheme \cite{buchmann2016high}, with logic locking techniques. Our solution focuses on securing the privilege switching process in RISC-V-based SoC designs, implementing encryption operations in critical logic paths while maintaining system functionality.

We implement and evaluate our approach on a RISC-V SoC based on the Rocket-Chip generator \cite{asanovic2016rocket}, providing detailed analysis of area overhead, performance impact, and security implications.

The key contributions of this work include:
\begin{itemize}
    \item A logic locking scheme that prevents parameter leakage through homomorphic encryption.
    \item An implementation of lattice-based cryptography in hardware security, providing potential quantum resistance.
    \item Experimental validation demonstrating the impact on the system performance.
\end{itemize}

The structure of this paper is as follows: Section 2 provides essential background information and a discussion of related work in the domains of logic locking and homomorphic encryption. Section 3 presents a detailed description of the proposed bin-RLWE-based logic locking scheme. Section 4 elucidates the experimental setup and analyzes the results obtained. Section 5 offers a comparison with related works and a discussion of the findings. Finally, Section 6 concludes the paper.

\IEEEpubidadjcol

\section{Background and Related Works}

\subsection{Logic Locking}

Logic locking is a hardware security technique designed to protect integrated circuit designs from unauthorized reproduction and tampering. The fundamental principle involves modifying the original circuit design through the insertion of specific locking structures, thereby ensuring that the circuit only functions correctly when provided with the appropriate key.

\begin{figure}[t]
    \centering
    \includegraphics[width=2.5in]{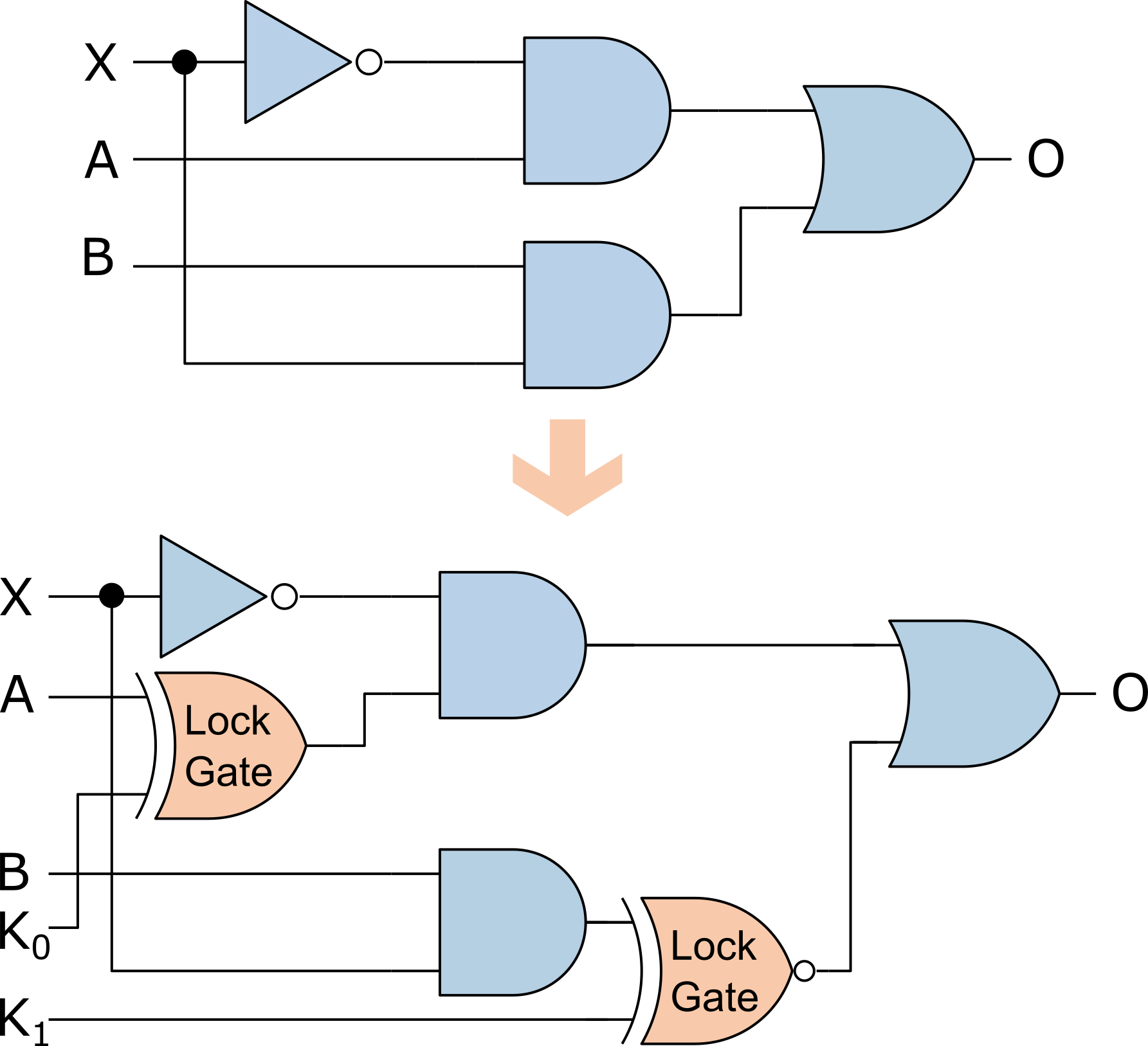}
    \caption{Basic example of combinational logic locking: a 2:1 selector with logic locking implemented through inserted XOR and XNOR gates, where the key is $K_0 = 0, K_1 = 1$.}
    \label{fig_lock_example}
\end{figure}

Logic locking typically entails the insertion of additional logic gates (e.g., XOR or XNOR) at strategic points within the circuit, rendering correct functionality dependent on a key input. Figure \ref{fig_lock_example} illustrates a basic implementation of logic locking, where $K_0 = 0, K_1 = 1$ represents the correct key. 

While these techniques have demonstrated promise, they face several significant challenges. The insertion of locking structures frequently results in increased silicon area and power consumption, particularly in complex circuits. Moreover, the development of attack methods, such as SAT-based attacks \cite{subramanyan2015evaluating} and side-channel analysis \cite{yasin2015security}, has exposed vulnerabilities in many logic locking schemes.

A critical concern with traditional logic locking is the potential for complete design recovery if the unlock key is compromised. This risk is particularly unacceptable in highly sensitive applications. Furthermore, the security of many logic locking schemes relies on the computational difficulty of solving SAT problems, which may be vulnerable to efficient solving algorithms in a quantum computing environment \cite{boulebnane2024solving,  tan2023hyqsat, alasow2022quantum}.

\subsection{Homomorphic Encryption}

Homomorphic encryption is a cryptographic paradigm that enables computations to be performed on encrypted data without requiring decryption, thus preserving data privacy during processing. The concept was initially proposed by Rivest, Adleman, and Dertouzos in 1978 \cite{rivest1978data}. However, it was not until 2009 that Craig Gentry introduced the first feasible fully homomorphic encryption (FHE) scheme based on ideal lattices \cite{gentry2009fully}.

Lattice-based homomorphic encryption schemes, particularly those based on the Ring Learning with Errors (RLWE) problem, have gained prominence due to their efficiency and potential resistance to quantum attacks. The bin-RLWE scheme, a variant of RLWE-based encryption, offers reduced computational and storage overhead through binarized noise and lower sampling depth \cite{buchmann2016high}.

The integration of homomorphic encryption techniques with hardware security measures presents a promising opportunity to address the vulnerabilities of traditional logic locking while leveraging the privacy-preserving properties of encrypted computation. This approach forms the foundation of our proposed bin-RLWE-based logic locking scheme, which aims to provide protection against both key leakage and potential quantum attacks.

\section{Proposed Method: Logic Locking Based on Homomorphic Encryption}

This study presents an approach that integrates homomorphic encryption technology with logic locking, applying it to the privilege switching process of SoC designs. This method enhances hardware security by introducing encryption operations in critical logic paths. The core concept leverages the unique characteristics of homomorphic encryption to perform operations on encrypted data without full decryption, thereby preserving the confidentiality of critical logic.

\subsection{Implementation Overview}

In our implementation, we identify and split a critical logic cone within the SoC's privilege switching logic to obtain intermediate input/output. We then selectively flip certain bits of the original intermediate output, with these flipped bits constituting the key $K$. The modified intermediate output is subsequently directed to an encryption path, which encodes and encrypts the data into the ciphertext domain.

\begin{figure}[b]
    \centering
    \includegraphics[width=\columnwidth]{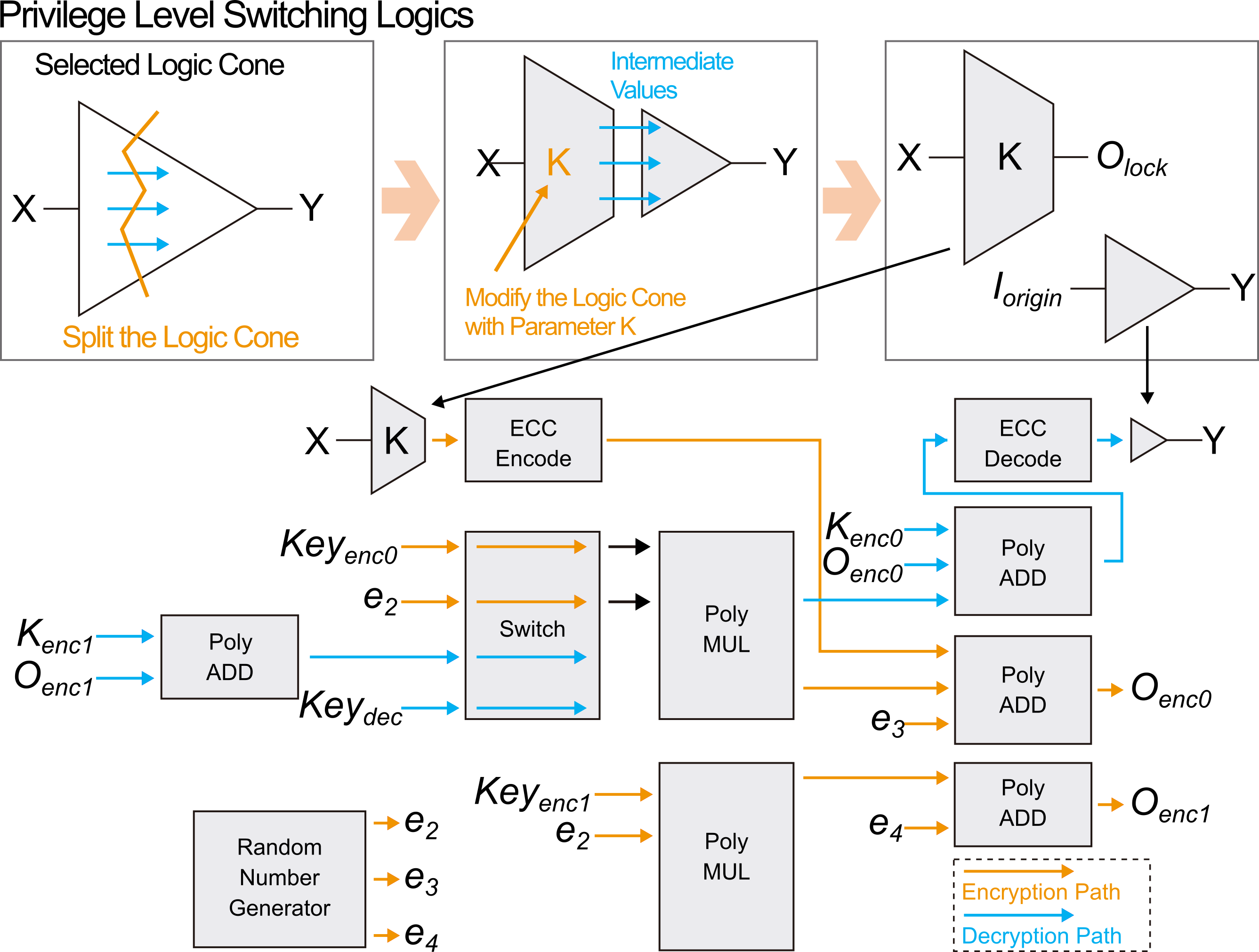}
    \caption{Proposed homomorphic encryption-based logic locking implementing procedure and architecture: the integration of the logic locking module is at a middle point of the processor privilege switching logic, with resource sharing between encryption and decryption paths.}
    \label{fig_sys_struct}
\end{figure}

The encrypted data is then transmitted to a decryption path, while a software control interface inputs the encrypted key $K_{enc}$ (the ciphertext domain representation of $K$) to the module. A homomorphic XOR operation is performed, combining $K_{enc}$ with the encrypted intermediate value. Finally, the decryption result yields the original intermediate input, which is then reconnected to the logic cone. Figure \ref{fig_sys_struct} illustrates the overall logic diagram of this process.

Notably, our implementation employs shift convolution for polynomial multiplication operations, achieving compact area utilization for smaller polynomial lengths. This approach establishes a linear correlation between latency and polynomial length.

\subsection{Homomorphic Encryption Algorithm}

We selected the bin-RLWE algorithm as the foundation for our homomorphic encryption scheme. To accommodate the constraints of hardware implementation, we have optimized the algorithm by modifying the parameters. We decreased the ciphertext length to minimize computation cycles and hardware resource consumption. The bit depth of the ciphertext domain is reduced to strike a balance between security and error rate.
A notable characteristic of bin-RLWE is its use of a binary distribution (0 and 1) for noise values. The noise strength is determined by the depth of the ciphertext domain, with shallower depths resulting in stronger noise, higher error rates, and increased security. While shortening the ciphertext length reduces the difficulty of potential attacks, it significantly decreases computation cycles and hardware overhead.

\subsection{Encryption and Decryption Process}

Algorithm \ref{algorithm_lock} delineates the overall procedure for our homomorphic encryption-based logic locking scheme. Following homomorphic computation, the result obtained is the original intermediate input with added noise. When the noise strength remains below a certain threshold, the decryption result maintains the correctness of the original logic.

Due to the reduction in the bit depth of the ciphertext domain, we observed an increase in the error rate. To mitigate this, our current implementation incorporates XOR check codes. We add Error-Correcting Codes (ECC) to the result after flipping the original output and verify the decryption result. If the verification fails, the encryption process is repeated. Future iterations will introduce more robust ECCs to enhance performance and reliability.

\begin{algorithm}[htbp]
\caption{Homomorphic Encryption-based Logic Locking}
\label{algorithm_lock}
\begin{algorithmic}
\STATE \textbf{Parameters:}
\STATE \hspace{0.5cm} $n$: Number of bits in input and output
\STATE \hspace{0.5cm} $q$: Modulus of the ring
\STATE \textbf{Input:}
\STATE \hspace{0.5cm} $K$: Locking parameter determined by the chip designer
\STATE \hspace{0.5cm} $O_{origin}$: Original intermediate output
\STATE \textbf{Output:}
\STATE \hspace{0.5cm} $I_{origin}$: Original intermediate input
\STATE \textbf{Variables:}
\STATE \hspace{0.5cm} $Key_{dec}$ : Decryption key
\STATE \hspace{0.5cm} $Key_{enc0,1}$ : First and second part of the encryption key
\STATE \hspace{0.5cm} $O_{lock}$:  Locked intermediate output
\STATE \hspace{0.5cm} $O_{enc}$:  Locked and encrypted intermediate output
\STATE \hspace{0.5cm} $K_{enc}$:  Encrypted locking parameter
\STATE \hspace{0.5cm} $T_{mid}$:  Intermediate temporary variable
\STATE \hspace{0.5cm} $e_i \stackrel{\$}{\leftarrow} \{0,1\}^n$, for $i \in \{0,1,2,3,4,5,6,7\}$
\STATE \textbf{Functions:}
\STATE \hspace{0.5cm} $\textsc{Extend}$: Extend from binary to ring
\STATE \hspace{0.5cm} $\textsc{Reduce}$: Reduce from ring to binary
\STATE {\textsc{Key Pair Generation}} 
\STATE \hspace{0.5cm} $Key_{dec} \gets e_0$ \hfill (1)
\STATE \hspace{0.5cm} $a_1 \stackrel{\$}{\leftarrow} \mathbb{Z}_q^n$ \hfill (2)
\STATE \hspace{0.5cm} $a_0 \gets - (e_1 + a_1 \cdot Key_{dec})$ \hfill (3)
\STATE \hspace{0.5cm} $(Key_{enc0}, Key_{enc1}) \gets (a_0, a_1)$ \hfill (4)
\STATE {\textsc{Encryption}}
\STATE \hspace{0.5cm} $O_{lock} \gets \textsc{Extend}(O_{origin} \oplus K)$ \hfill (5)
\STATE \hspace{0.5cm} $O_{enc_0} \gets O_{lock} + Key_{enc0} \cdot e_2 + e_3$ \hfill (6)
\STATE \hspace{0.5cm} $O_{enc_1} \gets Key_{enc1} \cdot e_2 + e_4$ \hfill (7)
\STATE {\textsc{Locking Parameter Encryption}} 
\STATE \hspace{0.5cm} $K_{enc_0} \gets \textsc{Extend}(K) + Key_{enc0} \cdot e_5 + e_6$ \hfill (8)
\STATE \hspace{0.5cm} $K_{enc_1} \gets Key_{enc1} \cdot e_5 + e_7$ \hfill (9)
\STATE {\textsc{Decryption}} 
\STATE \hspace{0.5cm} $T_{mid_0} \gets O_{enc_0} + K_{enc_0}$ \hfill (10)
\STATE \hspace{0.5cm} $T_{mid_1} \gets O_{enc_1} + K_{enc_1}$ \hfill (11)
\STATE \hspace{0.5cm} $O_{dec} \gets T_{mid_0} + T_{mid_1} \cdot Key_{dec}$ \hfill (12)
\STATE \hspace{0.5cm} $I_{origin} \gets \textsc{Reduce}(O_{dec})$ \hfill (13)
\STATE {\textsc{Output}}
\STATE \hspace{0.5cm} $O_{dec} = T_{mid_0} + T_{mid_1} \cdot Key_{dec}$
\STATE \hspace{0.5cm} $\phantom{O_{dec}} = (O_{enc_0} + K_{enc_0}) + (O_{enc_1} + K_{enc_1}) \cdot Key_{dec}$
\STATE \hspace{0.5cm} $\phantom{O_{dec}} = (O_{lock} + a_0 \cdot e_2 + e_3 + \textsc{Extend}(K) + a_0 \cdot e_5 + e_6) + (a_1 \cdot e_2 + e_4 + a_1 \cdot e_5 + e_7) \cdot e_0$
\STATE \hspace{0.5cm} $\phantom{O_{dec}} = (O_{lock} + \textsc{Extend}(K)) - (e_2 + e_5) \cdot e_1 + (e_4 + e_7) \cdot e_0 + e_3 + e_6$
\STATE 
\STATE \hspace{0.5cm} $\textsc{Noise} = - (e_2 + e_5) \cdot e_1 + (e_4 + e_7) \cdot e_0 + e_3 + e_6$
\STATE 
\STATE \hspace{0.5cm} $\textsc{Reduce}(O_{lock} + \textsc{Extend}(K)) = O_{origin} \oplus K \oplus K$
\STATE \hspace{0.5cm} $\phantom{\textsc{Reduce}(O_{lock} + \textsc{Extend}(K))} = O_{origin}$
\end{algorithmic}
\end{algorithm}

\subsection{Key Management and Security Features}

In our implementation of Algorithm \ref{algorithm_lock}, we employ a two-part approach for $Key_{dec}$: One part is stored in tamper-proof memory burned into the chip by the provider, while the other part is generated by hardware at power-up to prevent replay attacks. The use of tamper-proof memory prevents physical attacks from extracting $Key_{dec}$, which could otherwise enable attackers to decrypt $K_{enc}$.
Upon power-up, the system queries the key provider, supplying the chip number and a power-up-specific random number. The key provider retrieves the corresponding pre-burned content using the chip number and combines it with the random number to create $Key_{dec}$. Subsequently, $Key_{enc}$ is derived from $Key_{dec}$, and the locking parameter is encrypted using $Key_{enc}$ before being returned to the chip.

During the decryption process, the chip only needs to provide the encrypted locking parameter post-locking, thereby preventing leakage of the original encrypted parameter. The implementation of step (5) in Algorithm \ref{algorithm_lock} is achieved by directly modifying the chip's logic netlist. While we position the addition of the locking parameter $K$ before the encryption and decryption process in our current implementation, alternative placements—either after the process or simultaneously before and after—are also feasible.

\subsection{Hardware and Software Integration}

For the hardware implementation, we have developed a Pseudo-Random Number Generator (PRNG) based on Linear Feedback Shift Registers (LFSR) for homomorphic encryption. 

On the software side, we have modified the Linux kernel's interrupt and exception handling code to interact with the logic locking module at the entry point. This interaction includes key writing and waiting for the privilege level switch to complete. These modifications were necessary to accommodate the additional clock cycles required for the encryption and decryption process, which would otherwise disrupt the standard operation of the Linux kernel.

In our modified system, when an interrupt occurs, the processor maintains a low privilege level upon entering the interrupt handling function. The system only transitions to a high privilege level after the key is written via the logic lock handler and the related logic operations are completed.

\begin{figure}[b]
    \centering
    \includegraphics[width=3.3in]{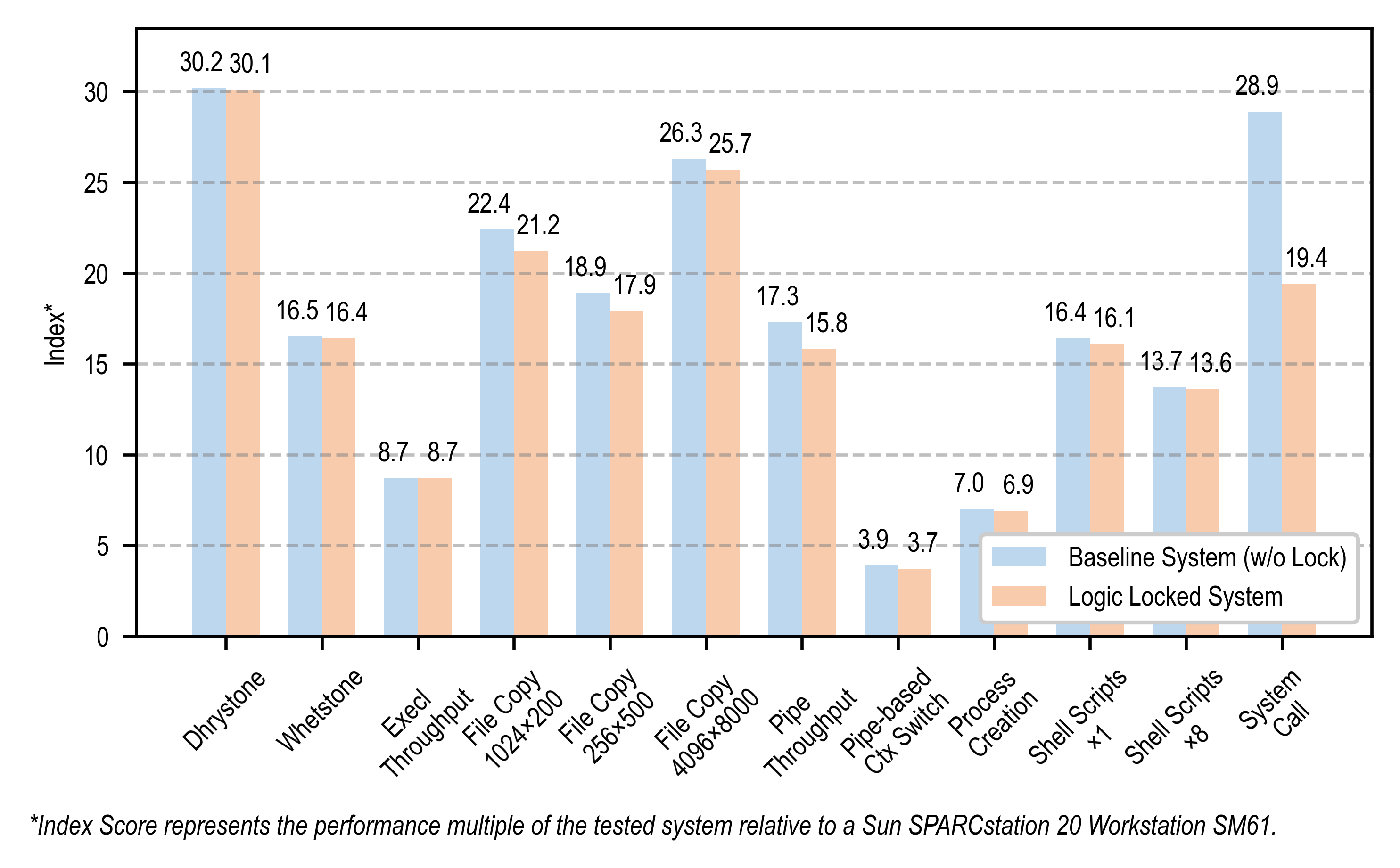}
    \caption{Performance comparison of Unixbench: baseline vs. locked implementation.}
    \label{fig_benchmark_result}
\end{figure}

\section{Experimental Design and Results}

\subsection{Experimental Setup}

\begin{figure}[b]
    \centering
    \includegraphics[width=3.1in]{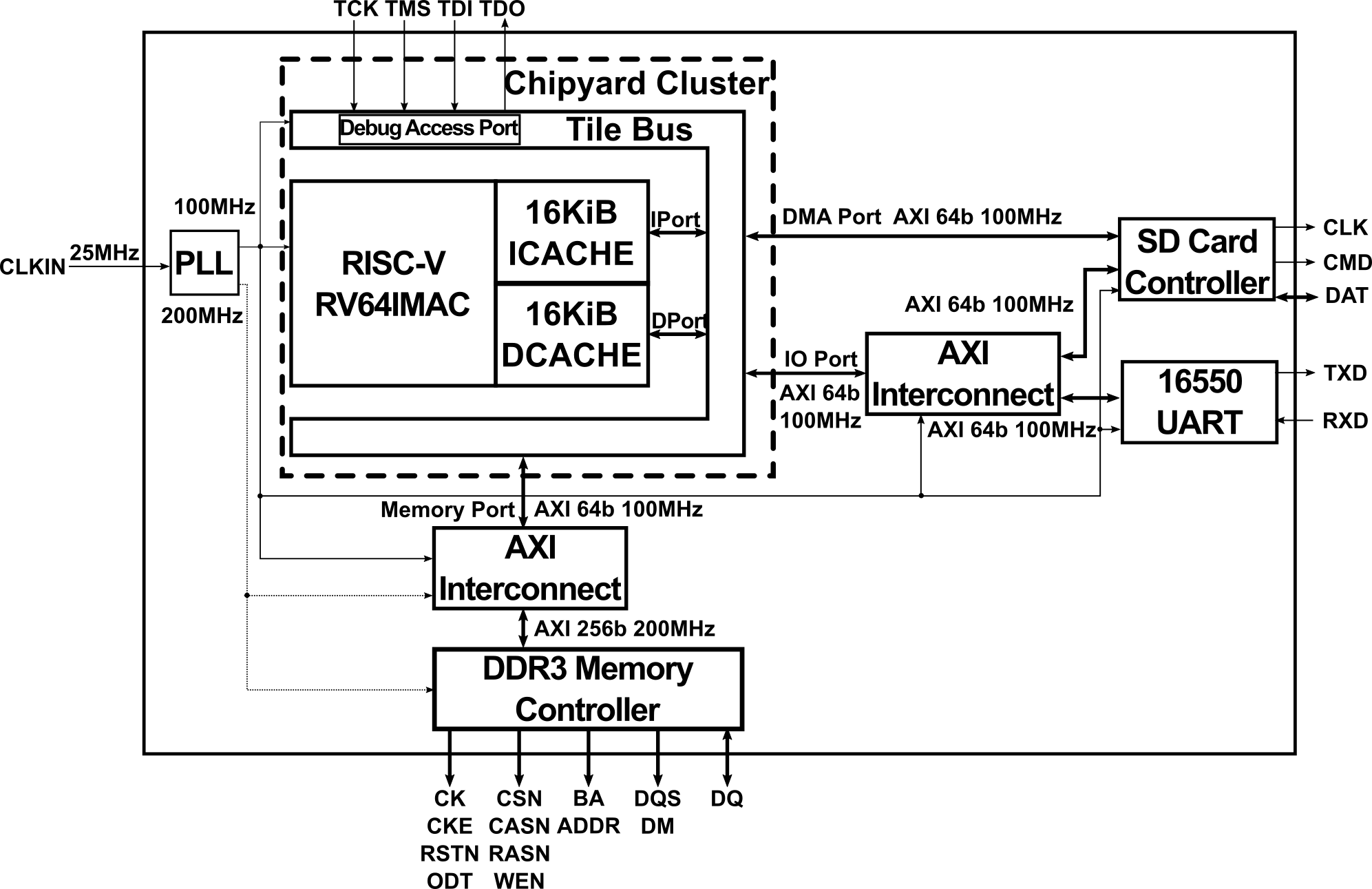}
    \caption{Block Diagram of SoC Architecture.}
    \label{sysfigure}
\end{figure}

Our experimental platform utilized Xilinx's XC7K160T FPGA for implementing and testing our proposed logic locking method. The SoC design operated within a GNU/Linux environment, specifically kernel version 6.1.6, with GCC version 14.2.0 serving as the compilation environment. The SoC design based on RocketChip 1.6.0 \cite{asanovic2016rocket} and Chipyard 1.11.0 \cite{chipyard} was shown in Figure \ref{sysfigure}.

\subsection{Performance Evaluation Methodology}

To assess the impact of our proposed method on system performance, we employed the BYTE UNIX Benchmarks (Version 5.1.3) \cite{unixbench} as our evaluation tool. This benchmark suite encompasses a series of system-level performance tests, offering a comprehensive reflection of various aspects of operating system performance.

We used the benchmark's Index Score as our performance indicator, which measures performance relative to a baseline Sun SPARCstation 20 Workstation SM61.

\subsection{Performance Results and Analysis}

The experimental results, as illustrated in Figure \ref{fig_benchmark_result}, revealed several noteworthy performance changes:

1. System Call Performance: We observed a significant decrease in system call performance, primarily attributed to the additional encryption and decryption processes triggered by privilege level switching during system calls. Specifically, the System Call Overhead Index decreased from 28.9 to 19.4, representing a substantial 32.9\% reduction.

2. File Operations: Operations involving privilege level switching, such as file copying, exhibited varying degrees of performance decline. Depending on the specific file copying pattern, performance decreased to between 94.6\% and 97.7\% of the original speed.

3. Pipe Operations: Pipe Throughput decreased to 91.3\% of its original value, indicating a moderate impact on inter-process communication efficiency.

4. Computational Performance: Notably, general computational performance, as indicated by Dhrystone and Whetstone test results, remained largely unaffected by our proposed method. This observation suggests that our approach primarily impacts system-level operations while having minimal effect on user-level computational tasks.

\subsection{Security Analysis}

The security of our method is based on the bin-RLWE algorithm, which employs binary noise instead of Gaussian noise. This approach renders our method average-case safe rather than worst-case safe \cite{buchmann2016hardness}. The original bin-RLWE parameters (n=256, bitdepth=8 \& q=256) provide a security level of 84 bits \cite{buchmann2016high}. According to the primal attack estimator in \cite{albrecht2015concrete}, our current implementation—employing parameters of n=54, bitdepth=7 \& q=128 —results in a diminished security level of 41 bits. However, it is important to note that the primal attack estimator assumes an infinite number of $Key_{enc}$ samples for a single $Key_{dec}$. When the number of samples is limited, the attack difficulty increases significantly \cite{bindel2019}. Since our scheme generates $Key_{dec}$ at each power-up, and for each corresponding $Key_{dec}$ only one $Key_{enc}$ is generated, from an LWE perspective, it is impossible to obtain an exact solution.

A key advantage of our method over traditional logic locking approaches is that users can only access ciphertext-domain locking parameters, with the original locking parameters never exposed. This characteristic significantly enhances system security by eliminating the possibility of users recovering the original design through locking parameters leakage during tranfer. Furthermore, the lattice-based cryptography underlying bin-RLWE potentially offers resistance against quantum attacks, a feature of increasing importance as we approach the era of quantum computing.

To evaluate its resistance against SAT attacks, we locked an n-bits XOR gate using this method while keeping the random numbers fixed, and attempted to solve for $K_{enc}$ using the attack method in \cite{subramanyan2015evaluating}. Our experimental results indicate that the computational complexity generally follows $2.5^{n \cdot bitdepth}$. Even with fixed random numbers, the method demonstrates considerable resistance against conventional attacks.

\subsection{Resource Utilization}

The overhead introduced by logic locking is presented in Table \ref{tab_hardware_resource}. Due to the interaction cost with the locking module, the overhead attributed to logic locking does not precisely correspond to the module's resource consumption.

\begin{table}[!t]
\caption{Hardware Resource Consumption\label{tab_hardware_resource}}
\centering
\begin{tabular}{|c|c|c|c|}
\hline
\rowcolor[HTML]{BDD7EF}  & System w/o & System with  & Locking \\
\rowcolor[HTML]{BDD7EF}  & Locking    & Locking       & Module  \\ \hline
\cellcolor[HTML]{F8CBAD}LUTs  & 56526      & 59936 (+6.0\%) & 3519    \\ \hline
\cellcolor[HTML]{F8CBAD}Regs  & 35794      & 38262 (+6.9\%) & 2645    \\ \hline
\cellcolor[HTML]{F8CBAD}BRAMs & 25         & 25            & 0       \\ \hline
\cellcolor[HTML]{F8CBAD}DSPs  & 15         & 15            & 0       \\ \hline
\end{tabular}
\end{table}

\setlength{\tabcolsep}{1pt}

\begin{table}[!t]
\caption{Comparison with Logic Locking and RLWE Implementations\label{tab_comparison}}
\centering
\begin{tabular}{cccccc}
\hline
\rowcolor[HTML]{BDD7EF} 
\multicolumn{1}{|c|}{\cellcolor[HTML]{BDD7EF}}                     & \multicolumn{1}{c|}{\cellcolor[HTML]{BDD7EF}This} & \multicolumn{1}{c|}{\cellcolor[HTML]{BDD7EF}This*} & \multicolumn{1}{c|}{\cellcolor[HTML]{BDD7EF}{[}2{]}} & \multicolumn{1}{c|}{\cellcolor[HTML]{BDD7EF}{[}16{]}} & \multicolumn{1}{c|}{\cellcolor[HTML]{BDD7EF}{[}17{]}} \\ \hline
\multicolumn{1}{|c|}{\cellcolor[HTML]{F8CBAD}Area Overhead}        & \multicolumn{1}{c|}{6.0\%/6.9\%}                  & \multicolumn{1}{c|}{-}                             & \multicolumn{1}{c|}{1.43\%}                          & \multicolumn{1}{c|}{-}                                & \multicolumn{1}{c|}{-}                                \\ \hline
\multicolumn{1}{|c|}{\cellcolor[HTML]{F8CBAD}(n,q,$\sigma$)}           & \multicolumn{1}{c|}{(54,128,-)}                     & \multicolumn{1}{c|}{{\tightnum{(201,128,-)}}}                     & \multicolumn{1}{c|}{-}                               & \multicolumn{1}{c|}{\tightnum{(256,4096,8.35)}}                  & \multicolumn{1}{c|}{\tightnum{(256,7681,11.31)}}                 \\ \hline
\multicolumn{1}{|c|}{\cellcolor[HTML]{F8CBAD}LUTs}                 & \multicolumn{1}{c|}{3519}                         & \multicolumn{1}{c|}{13497}                         & \multicolumn{1}{c|}{-}                               & \multicolumn{1}{c|}{1974+1698}                        & \multicolumn{1}{c|}{1381}                             \\ \hline
\multicolumn{1}{|c|}{\cellcolor[HTML]{F8CBAD}Regs}                 & \multicolumn{1}{c|}{2645}                         & \multicolumn{1}{c|}{10077}                         & \multicolumn{1}{c|}{-}                               & \multicolumn{1}{c|}{2698+1958}                        & \multicolumn{1}{c|}{1179}                             \\ \hline
\multicolumn{1}{|c|}{\cellcolor[HTML]{F8CBAD}Latency}              & \multicolumn{1}{c|}{2.6\textmu s}                        & \multicolumn{1}{c|}{2.5\textmu s}                         & \multicolumn{1}{c|}{-}                               & \multicolumn{1}{c|}{47\textmu s}                             & \multicolumn{1}{c|}{193\textmu s}                            \\ \hline
\multicolumn{1}{|c|}{\cellcolor[HTML]{F8CBAD}\tightword{Original Design Safe}} & \multicolumn{1}{c|}{\checkmark}                            & \multicolumn{1}{c|}{-}                             & \multicolumn{1}{c|}{×}                               & \multicolumn{1}{c|}{-}                                & \multicolumn{1}{c|}{-}                                \\ \hline
\multicolumn{1}{|c|}{\cellcolor[HTML]{F8CBAD}Security Bits}        & \multicolumn{1}{c|}{41}                           & \multicolumn{1}{c|}{83}                            & \multicolumn{1}{c|}{-}                               & \multicolumn{1}{c|}{84}                               & \multicolumn{1}{c|}{81}                               \\ \hline
\multicolumn{6}{l}{\begin{tabular}[c]{@{}l@{}}*Independent locking module. Clock frequency: 164MHz.\\ Time for writing $K_{enc}$ into the locking module is excluded.\end{tabular}}    
\end{tabular}
\end{table}

\section{Comparison and Discussion}

We conducted comparative analyses between our implementation and existing logic locking and RLWE implementations. Table \ref{tab_comparison} presents this comparison. While our hardware overhead is relatively higher, to the best of our knowledge, this represents the first lattice-based cryptography implementation in logic locking. Notably, our approach maintains the security of the original design even in cases of decryption key leakage. Compared to other RLWE implementations, we achieved significantly reduced total encryption and decryption latency, minimizing the performance overhead when integrated into processors. For reference, the independently operated locking module demonstrates substantial latency advantage at an equivalent security level when communication overheads are excluded. It's important to note that while this work focuses on logic locking, it does not restrict speculative execution in processors. Attackers can still potentially leak critical data through cache side-channels by inducing mis-speculation of kernel code operations via user-mode programs.

\section{Conclusion}

This study presents a novel approach to hardware security by integrating lattice based homomorphic encryption with logic locking techniques in SoC designs. Our implementation on a RISC-V SoC demonstrates the feasibility of this method, offering enhanced protection against key leakage.

The experimental results reveal a modest performance impact, primarily affecting system-level operations while maintaining user-level computational performance. The hardware overhead, with increases of 6.0\% in LUTs and 6.9\% in Regs consumption, represents a reasonable trade-off for the enhanced protection of the original design.


\begin{thebibliography}{1}
\bibliographystyle{IEEEtran}

\bibitem{rajarathnam2020regds}
R. S. Rajarathnam, Y. Lin, Y. Jin, and D. Z. Pan, ``ReGDS: A Reverse Engineering Framework from GDSII to Gate-level Netlist,'' in \textit{2020 IEEE Int. Symp. Hardware Oriented Security and Trust (HOST)}, 2020, pp. 154--163.


\bibitem{rathor2024gatelock}
V. S. Rathor, M. Singh, K. S. Sahoo, and S. P. Mohanty, ``GateLock: Input-dependent key-based locked gates for SAT resistant logic locking,'' \textit{IEEE Trans. Very Large Scale Integr. (VLSI) Syst.}, vol. 32, no. 2, pp. 361--371, Feb. 2024.

\bibitem{boulebnane2024solving}
S. Boulebnane and A. Montanaro, ``Solving Boolean satisfiability problems with the quantum approximate optimization algorithm,'' \textit{PRX Quantum}, vol. 5, no. 3, pp. 030348, 2024.


\bibitem{buchmann2016high}
J. Buchmann, F. Göpfert, T. Güneysu, T. Oder, and T. Pöppelmann, ``High-performance and lightweight lattice-based public-key encryption,'' in \textit{Proc. 2nd ACM Int. Workshop IoT Privacy, Trust, and Security}, 2016, pp. 2--9.

\bibitem{asanovic2016rocket}
K. Asanovic, R. Avizienis, J. Bachrach, S. Beamer, D. Biancolin, C. Celio, H. Cook, D. Dabbelt, J. Hauser, A. Izraelevitz, \textit{et al.}, ``The rocket chip generator,'' \textit{EECS Dept., Univ. California, Berkeley, Tech. Rep. UCB/EECS-2016-17}, vol. 4, pp. 6--2, 2016.

\bibitem{subramanyan2015evaluating}
P. Subramanyan, S. Ray, and S. Malik, ``Evaluating the security of logic encryption algorithms,'' in \textit{2015 IEEE Int. Symp. Hardware Oriented Security and Trust (HOST)}, 2015, pp. 137--143.

\bibitem{yasin2015security}
M. Yasin, B. Mazumdar, S. S. Ali, and O. Sinanoglu, ``Security analysis of logic encryption against the most effective side-channel attack: DPA,'' in \textit{2015 IEEE Int. Symp. Defect and Fault Tolerance in VLSI and Nanotechnology Syst. (DFTS)}, 2015, pp. 97--102.

\bibitem{tan2023hyqsat}
S. Tan, M. Yu, A. Python, Y. Shang, T. Li, L. Lu, and J. Yin, ``HyQSAT: A hybrid approach for 3-SAT problems by integrating quantum annealer with CDCL,'' in \textit{2023 IEEE Int. Symp. High-Performance Comput. Architecture (HPCA)}, 2023, pp. 731--744.

\bibitem{alasow2022quantum}
A. Alasow and M. Perkowski, ``Quantum algorithm for maximum satisfiability,'' in \textit{2022 IEEE 52nd Int. Symp. Multiple-Valued Logic (ISMVL)}, 2022, pp. 27--34.

\bibitem{rivest1978data}
R. L. Rivest, L. Adleman, M. L. Dertouzos, \textit{et al.}, ``On data banks and privacy homomorphisms,'' \textit{Found. Secure Comput.}, vol. 4, no. 11, pp. 169--180, 1978.

\bibitem{gentry2009fully}
C. Gentry, ``Fully homomorphic encryption using ideal lattices,'' in \textit{Proc. 41st Annu. ACM Symp. Theory Comput.}, 2009, pp. 169--178.



\bibitem{chipyard}
UC Berkeley Architecture Research, ucb-bar/chipyard. (Oct. 1, 2024). Scala. Accessed: Oct. 2, 2024. [Online]. Available: https://github.com/ucb-bar/chipyard


\bibitem{unixbench}
K. Lucas, kdlucas/byte-unixbench. (Oct. 16, 2024). C. Accessed: Oct. 18, 2024. [Online]. Available: https://github.com/kdlucas/byte-unixbench


\bibitem{buchmann2016hardness}
J. Buchmann, F. Göpfert, R. Player, and T. Wunderer, ``On the hardness of LWE with binary error: Revisiting the hybrid lattice-reduction and meet-in-the-middle attack,'' in \textit{Proc. Int. Conf. Cryptol. Africa}, 2016, pp. 24--43.

\bibitem{albrecht2015concrete}
M. R. Albrecht, R. Player, and S. Scott, ``On the concrete hardness of Learning with Errors,'' \textit{Cryptology ePrint Archive, Paper 2015/046}, 2015.


\bibitem{yang2024lightweight}
Y. Yang, Z. Wang, J. Wang, J. Hou, Y. Su, and C. Yang, ``A lightweight and efficient encryption/decryption coprocessor for RLWE-based cryptography,'' \textit{IEEE Trans. Circuits Syst. II, Exp. Briefs}, early access, 2024.

\bibitem{zhang2020efficient}
Y. Zhang, C. Wang, D. E. S. Kundi, A. Khalid, M. O'Neill, and W. Liu, ``An efficient and parallel R-LWE cryptoprocessor,'' \textit{IEEE Trans. Circuits Syst. II, Exp. Briefs}, vol. 67, no. 5, pp. 886--890, May 2020.

\bibitem{bindel2019}
N. Bindel, J. Buchmann, F. Göpfert, and M. Schmidt, ``Estimation of the hardness of the learning with errors problem with a restricted number of samples,'' \textit{J. Math. Cryptol.}, vol. 13, no. 1, pp. 47--67, 2019.


\end{thebibliography}
\end{document}